\documentclass[
12pt,
reprint,
showpacs,preprintnumbers,
showkeys,
amsmath,amssymb,
aps,
prb,
]{revtex4-1}

\usepackage{graphicx}
\usepackage{dcolumn}
\usepackage{bm}
\usepackage{hyperref}
\usepackage{subfigure}
\usepackage{nicefrac}
\usepackage{color}
\usepackage{braket}
\usepackage{multirow}
\usepackage{verbatim}
\usepackage{bbold}
\usepackage{nicefrac}

\usepackage{epstopdf}
\usepackage{tikz}
\usepackage{gnuplottex}
\newcommand{\ii}{\mathrm{i}}

\usepackage[normalem]{ulem}

\begin{document}

\preprint{APS/123-QED}

\title{Resonant scattering due to adatoms in graphene: top, bridge, and hollow position}

\author{\surname{Susanne} Irmer}
\email[Emails to: ]{susanne.irmer@physik.uni-regensburg.de}
\author{\surname{Denis} Kochan}
\author{\surname{Jeongsu} Lee}
\author{\surname{Jaroslav} Fabian}

\affiliation{%
 Institute for Theoretical Physics, University of Regensburg,\\
 93040 Regensburg, Germany
 }%

\date{\today}

\begin{abstract}
  We present a theoretical study of resonance characteristics in graphene from adatoms with $s$ or $p_z$ character binding in top, bridge, and hollow positions. The adatoms are described by two tight-binding parameters: on-site energy and hybridization strength. We explore a wide range of different magnitudes of these parameters by employing $T$-matrix calculations in the single adatom limit and by tight-binding supercell calculations for dilute adatom coverage. We calculate the density of states and the momentum relaxation rate and extract the resonance level and resonance width. The top position with a large hybridization strength or, equivalently, small on-site energy, induces resonances close to zero energy. The bridge position, compared to top, is more sensitive to variation in the orbital tight-binding parameters. Resonances within the experimentally relevant energy window are found mainly for bridge adatoms with negative on-site energies. The effect of resonances from the top and bridge positions on the density of states and momentum relaxation rate is comparable and both positions give rise to a power-law decay of the resonant state in graphene. The hollow position with $s$ orbital character is affected from destructive interference, which is seen from the very narrow resonance peaks in the density of states and momentum relaxation rate. The resonant state shows no clear tendency to a power-law decay around the impurity and its magnitude decreases strongly with lowering the adatom content in the supercell calculations. This is in contrast to the top and bridge positions. We conclude our study with a comparison to models of pointlike vacancies and strong midgap scatterers. The latter model gives rise to significantly higher momentum relaxation rates than caused by single adatoms.
\end{abstract}

\maketitle

\section{\label{sec:intr}Introduction}
In the past decade, graphene research has made remarkable progress; from its first experimental characterization \cite{Novoselov2004}, the way was paved towards high-quality graphene devices \cite{Dean2010,Dean2011} and proximitized graphene as ingredients for electronic and spintronics applications \cite{Novoselov2012,Roche2015,Han2014,Gmitra:PRB2015,Avsar2017,Gurram2017}.

Graphene was the first realized two-dimensional crystal material with a linear dispersion at low energy. As the low-energy electrons can be described by an effective Dirac equation for massless fermions, graphene was suggested for studies of relativistic effects such as Klein tunneling or zitterbewegung \cite{Katsnelson2007}. Apart from this fundamental interest in the two-dimensional carbon allotrope, efforts were taken to tailor graphene properties for electronic and spintronics devices.

On the one hand, proximity effects in graphene were explored. It was found that exchange interaction can be induced in graphene by placing it on a ferromagnetic insulator \cite{Swartz2012} or, separated by a tunnel barrier, on a ferromagnetic metal \cite{Zollner:Prox2016}. Additionally, proximity-induced large spin-orbit coupling \cite{Frank2016,Gmitra2016} can cause topological effects \cite{Frank2017}, giant spin lifetime anisotropy \cite{Cummings2017,Ghiasi2017,Benitez2017}, and, together with proximity exchange, transport magnetoanisotropies \cite{Lee2016}. On the other hand, local adsorbates on graphene can be used to functionalize graphene. For example, graphene's intrinsic spin-orbit coupling---of the order of $10\,\mu$eV \cite{Gmitra2009}---was shown to be increased by more than a factor of 100  by adatoms, such as hydrogen and copper \cite{Neto2009,Gmitra:PRL2013,Frank:PRB2017} making the spin-Hall effect accessible in graphene \cite{Balakrishnan2013,Balakrishnan2014,Ferreira:PRL2014,Yang:PRB2016,Huang:PRB2016}. It was also predicted that neutral adatoms with large spin-orbit coupling may stabilize the quantum spin Hall state in graphene \cite{Weeks2011} although experimental challenges still remain \cite{Jia2015,Chandni2015,dosSantos2017}. Furthermore, unconventional transport regimes were reported in theoretical investigations of specific kinds of disorder \cite{Schelter2011,Gattenloehner2016}.

In experiments on orbital transport, long-range Coulomb scattering \cite{Adam2007,Swartz2013,Jia2015,Chandni2015} of charged adatoms can strongly affect the measurements, whereas short-range scattering off adatoms, on the other hand, highly influences spin relaxation \cite{McCreary2012,Wojtaszek2013,Kochan2014,Kochan2015,Raes2016,Omar2017,Lundeberg:PRL2013} as the electrons feel the adsorbate induced local spin-orbit coupling \cite{Neto2009,Gmitra:PRL2013,Irmer:PRB2015,Zollner:Meth2016,Frank:PRB2017} or magnetic moment. The local magnetic moments originate, for example, from $sp^3$ defects such as hydrogen adatoms \cite{Duplock2004,Yazyev2010,McCreary2012,Birkner2013}, organic molecules \cite{Santos2012,Zollner:Meth2016}, or vacancies \cite{Ugeda2010}.

Vacancies furthermore give rise to zero-energy states in graphene \cite{Pereira2006,Pereira2008,Nanda2012}. Due to the small density of states at low energy, graphene is especially sensitive to such induced states that affect strongly transport by resonant scattering \cite{Stauber2007,Ferreira:PRB2011,Monteverde2010,Robinson2008}. Another source for resonant states at low energy can be substitutional impurities \cite{Basko2008,Wehling2007,Pereira2008,Skrypnyk2006} or adsorbates in graphene. The latter have been studied by explicit tight-binding and density-functional theory calculations of specific adatoms \cite{Ihnatsenka:PRB2011,Wehling:PRB2010,Wehling:PRL2010,Wehling:PRB2009,Farjam2011,Gmitra:PRL2013,Zollner:Meth2016,Frank:PRB2017}. It was also realized by basic symmetry analysis that the adsorption position of an adatom plays an important role for the resonance scattering mechanism \cite{Ruiz2016,Uchoa2014,Weeks2011,Duffy2016}. For example, it was established that the $s$ orbital of an adatom in the hollow position is effectively decoupled from the states of graphene \cite{Ruiz2016} so that resonance scattering of such an orbital is strongly suppressed.

Here, we study resonant scattering off single adatoms on graphene for the three stable adsorption positions, namely, top, bridge, and hollow, within the $T$-matrix formalism. We concentrate on adatoms with $s$ or $p_z$ orbital character and characterize them by hybridization strength and on-site energy in a minimal tight-binding model. Our work extends and connects previous theoretical studies that are available in the relevant literature on this topic. For example, Wehling \textit{et al.}\onlinecite{Wehling2007} studied within the $T$-matrix formalism single and double substitutional impurities. They showed the impact of selected orbital parameters on the local density of states around the impurity and addressed also the case of magnetic impurities. Here, we do not consider substitutional impurities but rather adsorbed elements which alter, for example, the energy dependence of the local density of states. In contrast, Robinson \textit{et al.}~\onlinecite{Robinson2008} studied H$^+$ and OH$^-$ adsorbing in the top position. For small impurity concentrations they employed the $T$-matrix formalism and showed the rise of an asymmetry in the conductivity due to the adsorbate in contrast to the symmetric contribution of localized charged scatterers. We calculate resonance maps that scan a large portion of the orbital parameter space for resonance levels forming in the density of states and thus cover a broad variety of possible adsorbate realizations on graphene. These maps show that adatom induced peaks in the density of states are in bridge and hollow positions much more sensitive to the variation of orbital parameters than in top position. Furthermore, the density of states and momentum relaxation rate show that hollow adatoms are (almost) not hybridizing with the $\pi$ states of graphene. Considering the limit of dilute adatom concentration on graphene within supercell calculations, we investigate the localization of the resonant states and find a clear power-law decay for top and bridge adatoms in contrast to hollow adatoms. We complement our resonance analysis by a comparison of induced resonances from general adatoms with vacancies and the model of strong midgap scatterers \cite{Peres:PRB2006,Ostrovsky:PRB2006,Ferreira:PRB2011} in graphene. The work of Ferreira \textit{et al.} \onlinecite{Ferreira:PRB2011} assumes a strong resonant scatterer sitting in the top position on graphene. In their study of conductivity in single-layer and (biased) bilayer graphene they stress that the first Born approximation is not valid for strong resonant scatterers. Using partial wave analysis, they derive under certain approximations an analytic formula for the influence of strong resonant scatterers on the conductivity. We comment later in the manuscript on the applicability of their assumptions and stress the consequences for quantitative analyses based on this formula.

The paper is organized as follows.
We introduce in Sec.~\ref{sec:formal} the framework of $T$-matrix formalism and investigated adatom models.
Sections~\ref{sec:top}, \ref{sec:bridge}, and \ref{sec:hollow} present our resonance analysis for the top, bridge, and hollow adsorption position, respectively.
The localization of the resonant states is discussed in Sec.~\ref{sec:local}, followed by a comparison between adatoms and vacancies in Sec.~\ref{sec:vac}, before we conclude in Sec.~\ref{sec:sum}.

\section{\label{sec:formal}Method}
\subsection{\label{sec:tmat}$T$-matrix formalism}
We study resonances from monovalent adatoms on graphene in the single adatom limit within the non-perturbative $T$-matrix approach.
Given a system described by the Hamiltonian $\mathcal{H}=\mathcal{H}_0 + \mathcal{V}$, with $\mathcal{V}$ being the perturbation to the unperturbed system $\mathcal{H}_0$, the retarded Green's operator satisfies
\begin{equation}
\left[ E^+ - \mathcal{H} \right]\, \mathcal{G}(E^+) = \mathbb{1}\:,\label{Eq:greensDef}
\end{equation}
where $E^+ = E + \ii \delta$ and $\delta \rightarrow 0$ is an infinitesimal imaginary part.
From the Dyson equation, the full retarded Green's operator is given by
\begin{equation}
\mathcal{G}(E^+) = \mathcal{G}_0(E^+) + \mathcal{G}_0(E^+)\mathcal{T}\mathcal{G}_0(E^+)\:,
\end{equation}
with $\mathcal{T}=\mathcal{V}\left[ \mathbb{1} - \mathcal{G}_0(E^+) \mathcal{V} \right]^{-1}$ being the $T$-matrix.
In the case of a single adatom on graphene we use the standard nearest-neighbor tight-binding Hamiltonian of graphene $\mathcal{H}_0$ (hopping strength $t=2.6\,\mathrm{eV}$),
\begin{equation}\label{Eq:GrapheneH}
\mathcal{H}_0 = -t \sum\limits_{\langle l,m \rangle} \ket{c_l}\bra{c_m}\:,
\end{equation}
whereas the presence of the adatom enters as a local (energy-dependent) perturbation $\mathcal{V}$.
We obtain this perturbation by integrating out the adatom from the system by the L\"{o}wdin transformation (see Sections \ref{sec:adModels} and \ref{sec:results} for details).\\
From the $T$-matrix and the Green's functions of the unperturbed graphene (see Appendix \ref{app:green}), we directly obtain the perturbed density of states (DOS) per atom $\nu(E)=\nu_0(E)+\Delta\nu(E)$, where $\nu_0(E)\approx |E|/W^2$ is the unperturbed DOS per atom with $W=\sqrt{\sqrt{3}\pi}t$, and
\begin{equation}
\Delta\nu(E) = \frac{\eta}{\pi}\mathrm{Im}\,\mathrm{Tr}\left[ \frac{\partial}{\partial E}\mathcal{G}_0(E^+)\mathcal{T}(E^+) \right],
\end{equation}
is the correction to the DOS of pristine graphene introduced by the impurity \cite{Kochan:Proc2015}, where $\eta=1/(2N)$ is the impurity concentration and $N$ is the number of unit cells in graphene.\\
The itinerant electrons resonantly scattering off the adatom form the virtual bound state inducing a peak in the DOS \cite{Hewson1993}. At the energy of the peak position, the resonance energy, the wave function is power-law localized around the impurity \cite{Liang:PRL2012} due to the hybridization of the impurity level with graphene's low, though nonzero, DOS. The width (full width at half maximum) $\Gamma$ of the resonance represents the resonance life time $\tau = \hbar/\Gamma$. 

To characterize the resonances, we use the DOS, but we also show the resonant behavior in the momentum relaxation rate $\tau_m^{-1}$ (see Appendix \ref{app:momrel} for details). Note that the width of the resonance peaks in the momentum relaxation rate is in general different from the peak widths in the DOS.

\subsection{\label{sec:adModels}Adatom models}
\begin{figure}
\includegraphics[width=\columnwidth]{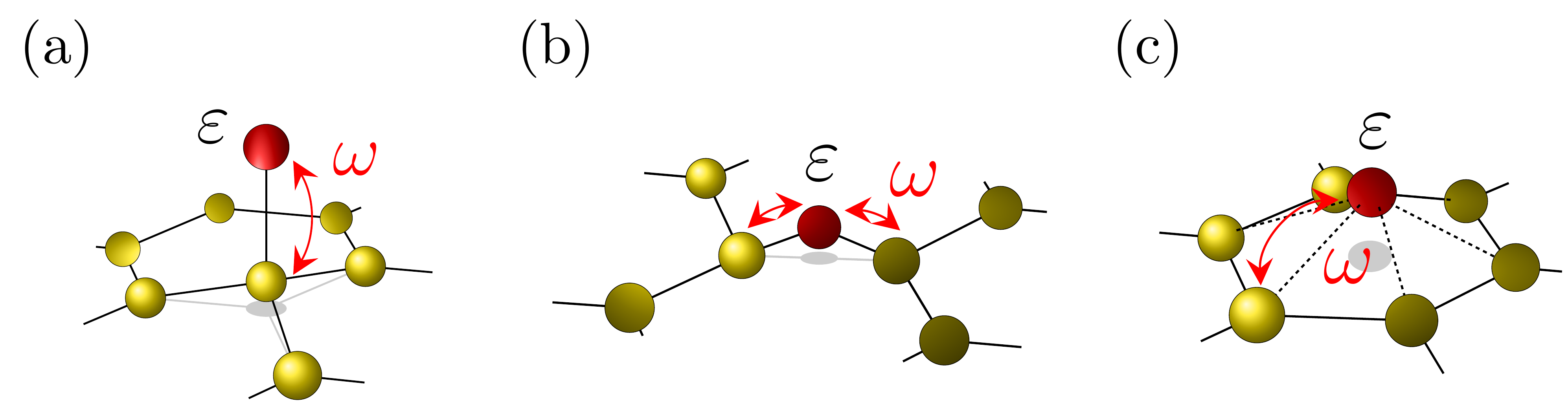}
\caption{\label{Fig:adatomsOrbitalSketch}(Color online) Sketch of the orbital hopping Hamiltonian for adatoms in (a) top, (b) bridge, and (c) hollow position. The adatom is modeled by on-site energy $\varepsilon$ and hybridization $\omega$ which connects the adatom to its nearest neighbors in graphene. }
\end{figure}
We describe the adatom on graphene in the single-electron picture by an on-site energy $\varepsilon$ of a single adatom orbital $\ket{X}$ and its hybridization $\omega$ to the nearest carbon neighbors in graphene.
The orbital is assumed to be invariant under $C_{6v}$ point group, so it has $s$ or $p_z$ orbital character. This kind of model has already been used successfully in tight-binding investigations based on first-principles calculations for the top and bridge positions \cite{Gmitra:PRL2013,Irmer:PRB2015,Zollner:Meth2016,Frank:PRB2017}.

Three stable adsorption positions are typical for adatoms on graphene: top, bridge, and hollow. As depicted in Fig.~\ref{Fig:adatomsOrbitalSketch}, the adsorption positions defer by the number of carbon hybridization partners for the adatom. While in the top position only one carbon atom contributes to the adsorption bond, bridge and hollow positions offer two and six bonding partners, respectively, for the adatom. The system is described by the Hamiltonian
\begin{equation}
H=\varepsilon\ket{X}\bra{X}+\omega\sum\limits_{\left\langle X, l\right\rangle}\left(\ket{X}\bra{c_l}+\mathrm{H.c.}\right)+\mathcal{H}_0\:,\label{Eq:GeneralAdatomH}
\end{equation}
where $l$ counts one, two, or six carbon sites, depending on the adsorption position, and $\mathcal{H}_0$ is the pristine graphene Hamiltonian, Eq.~(\ref{Eq:GrapheneH}).

By down-folding $H$, we obtain the Hamiltonian $\mathcal{H} = \mathcal{H}_0 + \mathcal{V}$, including only graphene degrees of freedom, and an energy-dependent perturbation acting on the nearest carbon neighbor(s) of the adatom,
\begin{align}
\mathcal{V}(E) &= \frac{|\omega|^2}{E-\varepsilon} \:\mathcal{P}\:,\label{Eq:GeneralAdatomPert}
\end{align}
where $E$ is the energy and $\mathcal{P}$ is the projection to the space of states formed by the $p_z$ orbitals of the hybridization partners of the adatom,
\begin{equation}
\mathcal{P} = \left(\sum\limits_{\left\langle X, l\right\rangle} \ket{c_l} \right) \left(\sum\limits_{\left\langle X, m\right\rangle} \bra{c_m} \right)\,.\label{Eq:ProjectionFactor}
\end{equation}
The hybridization partners are therefore coupled among themselves by $\mathcal{V}(E)$.
The non-vanishing block of the $T$-matrix can be written as
\begin{equation}
\tilde{\mathcal{T}} = \frac{|\omega|^2}{E-\varepsilon - \omega^2 \mathcal{A}(E)} \:\mathcal{P}\:,\label{Eq:tmatGeneral}
\end{equation} 
where $\mathcal{A}$ describes a combination of (retarded) Green's functions that depend on the adsorption position,
\begin{equation}\label{Eq:A}
\mathcal{A}=\begin{cases}G_{00}(E) & \mbox{(top)}\\[0.5em]
2\left[G_{00}(E)+G^\mathrm{AB}_{12}(E)\right]& \mbox{(bridge)}\\[0.5em]
6 \left\{ G_{00}(E) + G^{\rm AB}_{14}(E) \right. & \\\phantom{6\{}+ \left. 2 [G^{\rm AB}_{12}(E) + G^{\rm NN}_{13}(E)] \right\} &\mbox{(hollow)}
\end{cases}
\end{equation}
Here, $G_{00}(E)$ is the (retarded) on-site Green's function, $G^\mathrm{AB}_{12}$ and  $G^{\rm AB}_{14}$ are the nearest and third-nearest neighbor Green's functions of unperturbed graphene, respectively, which naturally couple opposite sublattices. Second-nearest neighbors on the same sublattice are coupled by $G^{\rm NN}_{13}$ (see Appendix~\ref{app:green} for details).

\section{\label{sec:results}Results and discussion}

\subsection{\label{sec:top}Top position}
\begin{figure}
\includegraphics[width=\columnwidth]{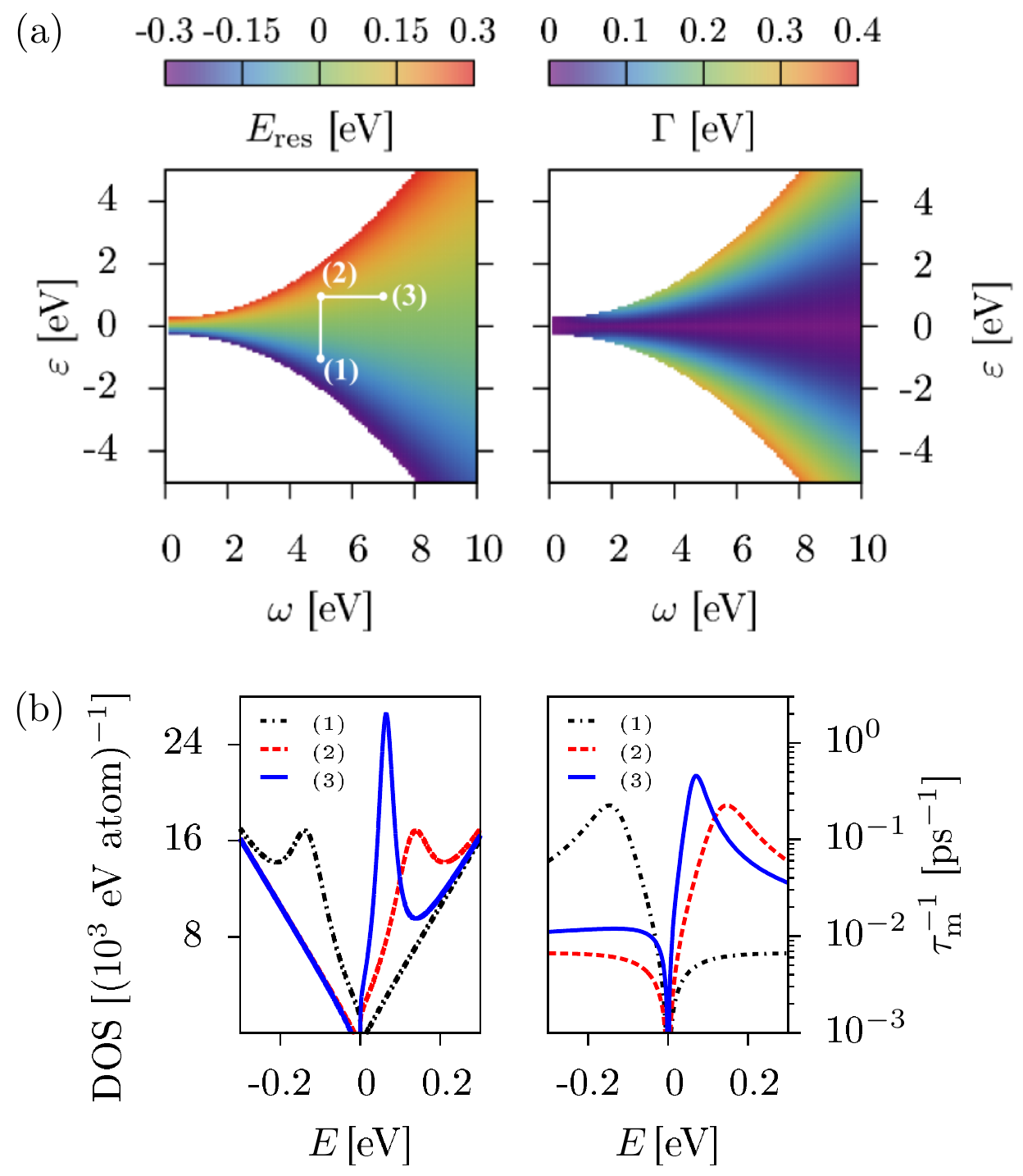}
\caption{\label{Fig:topResonanceMap_DosMomrel}(Color online) Resonance and momentum relaxation characteristics due to adatoms in top position. (a) Resonance energy $E_\mathrm{res}$ and width $\Gamma$ are shown as functions of $\varepsilon$ and $\omega$. (b) Snapshots of DOS and $\tau_m^{-1}$ at three parameter sets (1) $\omega = 5\,$eV and $\varepsilon = -1\,$eV, (2) $\omega = 5\,$eV and $\varepsilon = 1\,$eV, and (3) $\omega = 7\,$eV and $\varepsilon = 1\,$eV. The resonance levels are, respectively, at (1) $E_\mathrm{res}=-130\,$meV with $\Gamma=109\,$meV, (2) $E_\mathrm{res}=130\,$meV with $\Gamma=109\,$meV, and (3) $E_\mathrm{res}=64\,$meV with $\Gamma=50\,$meV. DOS data are shown for adatom concentration of $\eta = 10^{3}\,$ppm, for better resolution, and momentum relaxation rates for realistic $\eta=1\,$ppm.}
\end{figure}

Following the procedure of Sec.~\ref{sec:formal}, we extract the resonance energy $E_\mathrm{res}$ and width $\Gamma$ from the DOS under variation of orbital parameters $\omega$ and $\varepsilon$ and show DOS and $\tau_m^{-1}$ for specific parameters, see Fig.~\ref{Fig:topResonanceMap_DosMomrel}. We restrict ourselves to the experimentally relevant energy range of $E_{\rm res}$ in $[-0.3,0.3]$\,eV, which is equivalent to the variation of carrier density in the range $[-9.5,9.5]\,\mathrm{cm}^{-2}$.

If we lower $\omega$ for a fixed $\varepsilon$, we gradually decouple the adatom from graphene. At $\omega=0$, the isolated adatom level induces a $\delta$-peak on top of the DOS at $E=\varepsilon$.
In the top position the resonance energy is mainly determined by the singularity in the denominator of the $T$-matrix. Since the real part of the Green's function is an odd function of $E$ (the imaginary part is even), the resonance energy changes sign, $E_\mathrm{res}\rightarrow -E_\mathrm{res}$ for fixed $\omega$ and $\varepsilon\to -\varepsilon$. Apart from the sign in $E_{\rm res}$, the maps in Fig.~\ref{Fig:topResonanceMap_DosMomrel}(a) therefore exhibit mirror symmetry with respect to the $\varepsilon = 0$ axis.

Figure~\ref{Fig:topResonanceMap_DosMomrel}(b) presents DOS and momentum relaxation rate $\tau_m^{-1}$ for selected parameter sets, indicated by the path (1)-(2)-(3) in panel (a). Along path (1) to (2) the resonance level behaves as argumented above: two adatoms with same $\omega_1=\omega_2=\omega$ but $\varepsilon_1=-\varepsilon_2$ induce resonances on opposite sites of zero energy. Increasing the hybridization strength $\omega$ for fixed $\varepsilon$, path (2) to (3), the resonance level shifts closer to zero with decreasing width. In the limit of $\omega\to\infty$ we have the effective potential on the adsorption site, Eq.~\ref{Eq:GeneralAdatomPert}), $\omega^2/(E-\varepsilon)\to\infty$, which enforces the wave function to vanish there. This limit simulates a vacancy in graphene, which induces a zero-energy mode \cite{Ducastelle:PRB2013,Pereira2006,Peres:PRB2006,Pereira2008} (see also Sec.~\ref{sec:vac}). For a general impurity in top position we see that the larger the resonance energy $E_{\mathrm{res}}$ is the larger is the resonance width $\Gamma$. This is because with increased energy there are more graphene states to which the impurity level can couple to.

The momentum relaxation rate shows characteristic peaks at the resonance energies obtained from the DOS calculations with slightly different widths. The farther away from zero energy a resonance level forms the more noticeable is the electron-hole asymmetry \cite{Robinson2008} in the graphs of $\tau_m^{-1}$, Fig.~\ref{Fig:topResonanceMap_DosMomrel}(b). In experiments with dilute adatom coverage $\eta \simeq 1-100\,$ppm, for example fluorinated graphene of Ref.\onlinecite{Hong2011,Hong:PRL2012}, the asymmetry is most probably masked by the symmetric momentum relaxation rate profile of charged impurities in the substrate \cite{Sarma2011} or additional strong midgap scatterers \cite{Stauber2007} contributing to transport. Asymmetric transport behavior was in contrast observed in highly fluorinated graphene samples \cite{Tahara2013}.
 
\subsection{\label{sec:bridge}Bridge position}
\begin{figure}
\includegraphics[width=\columnwidth]{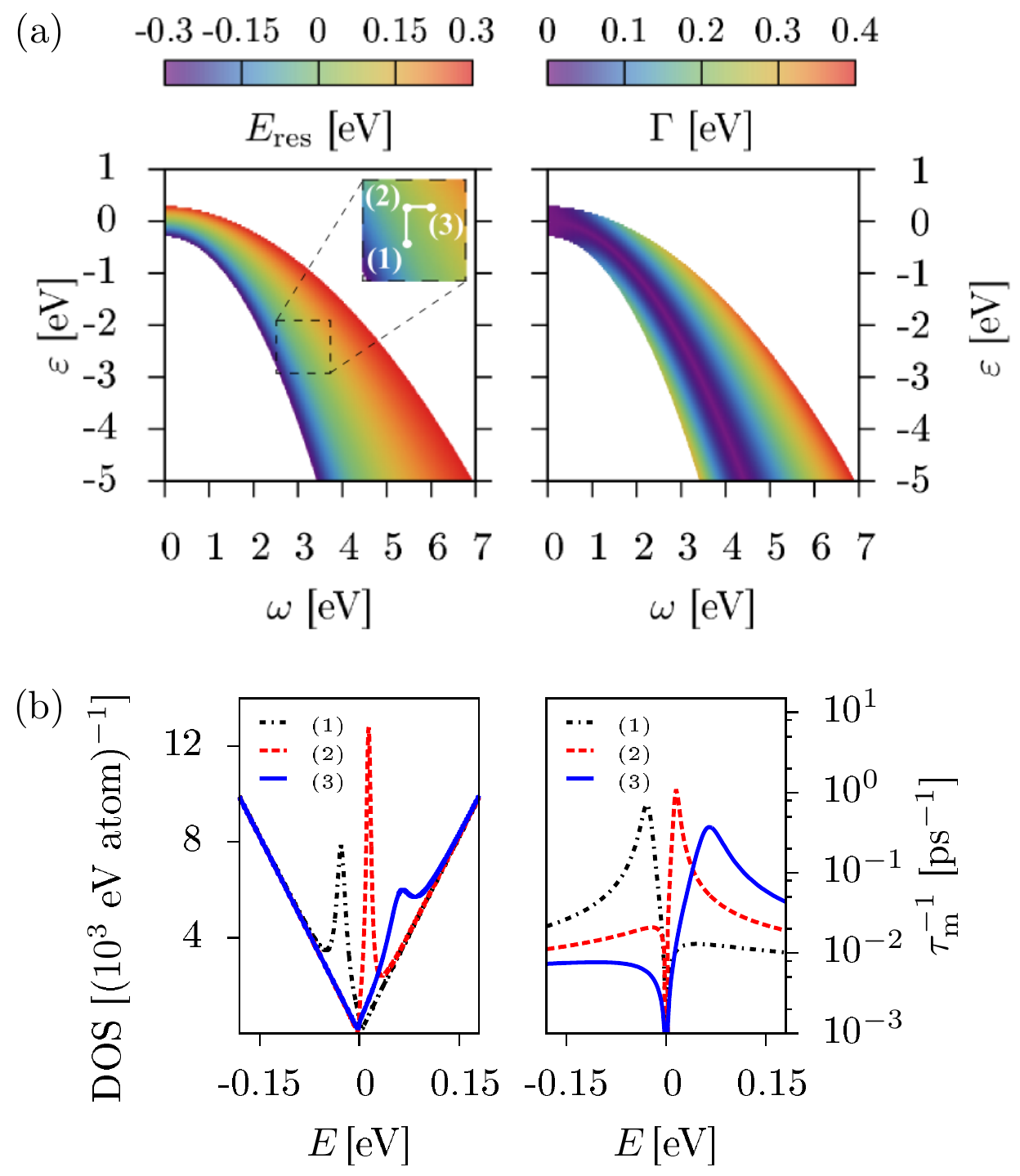}
\caption{\label{Fig:bridgeResonanceMap_DosMomrel}(Color online) Resonance and momentum relaxation characteristics due to adatoms in the bridge position. (a) Resonance energy $E_\mathrm{res}$ and width $\Gamma$ are shown as functions of $\varepsilon$ and $\omega$. (b) Snapshots of DOS and $\tau_m^{-1}$ at three parameter sets (1) $\omega = 3\,$eV and $\varepsilon = -2.5\,$eV, (2) $\omega = 3\,$eV and $\varepsilon = -2.2\,$eV, and (3) $\omega = 3.2\,$eV and $\varepsilon = -2.2\,$eV. The resonance levels are, respectively, at (1) $E_{\rm res}=-27\,$meV with $\Gamma=17\,$meV, (2) $E_{\rm res}=14\,$meV and $\Gamma=9\,$meV, and (3) $E_{\rm res}=60\,$meV and $\Gamma=42\,$meV. DOS data is shown for adatom concentration of $\eta = 100\,$ppm, for better resolution, and momentum relaxation rates for realistic $\eta=1\,$ppm.}
\end{figure}

The bridge position, Eq.~(\ref{Eq:A}), is affected by the interplay of Green's functions with different behavior under $E\to -E$ (see Appendix~\ref{app:green}). The symmetry arguments for resonance levels of the previous section are no longer valid. Figure~\ref{Fig:bridgeResonanceMap_DosMomrel}(a) shows $E_{\rm res}$ and $\Gamma$ extracted from the DOS for the bridge position. We observe that the parameter region leading to resonances, in the fixed energy interval $[-0.3,0.3]\,$eV, is dominated by negative on-site energy $\varepsilon$ and constrained to a smaller region compared to the top position. The resonance level position is much more sensitive to variation of parameters $\omega$ and $\varepsilon$ which was also observed in Ref.~\onlinecite{Wehling2007} for substitutional double impurities. Equating there the coupling strength $U_1$ between the substitutional impurities with their on-site potentials $U_0$ leads to a local perturbation comparable to Eq.~(\ref{Eq:GeneralAdatomPert}) for the bridge position. However, the perturbation describing  the substitutional double impurity is energy independent, whereas for a bridge adatom it is energy dependent due to the down-folding process.

Similar as in the case of a single top adatom we can shift the resonance level for fixed $\omega$ from negative to positive energies upon increasing $\varepsilon$ along path (1) to (2). The transition from negative to positive resonance energy is also visible in the DOS and $\tau_m^{-1}$ in Fig.~\ref{Fig:bridgeResonanceMap_DosMomrel}(b). Though, fixing the on-site energy and increasing the hybridization strength shifts the resonance further away from zero energy. This behavior is in contrast to the top position and originates from the effective coupling via the adatom between the two carbon hybridization partners in graphene.

Still we observe the natural broadening of peaks with increasing resonance energy. Furthermore, the magnitude of typical momentum-relaxation rate is comparable to the top position.

\subsection{\label{sec:hollow}Hollow position}

\begin{figure}
\includegraphics[width=0.95\columnwidth]{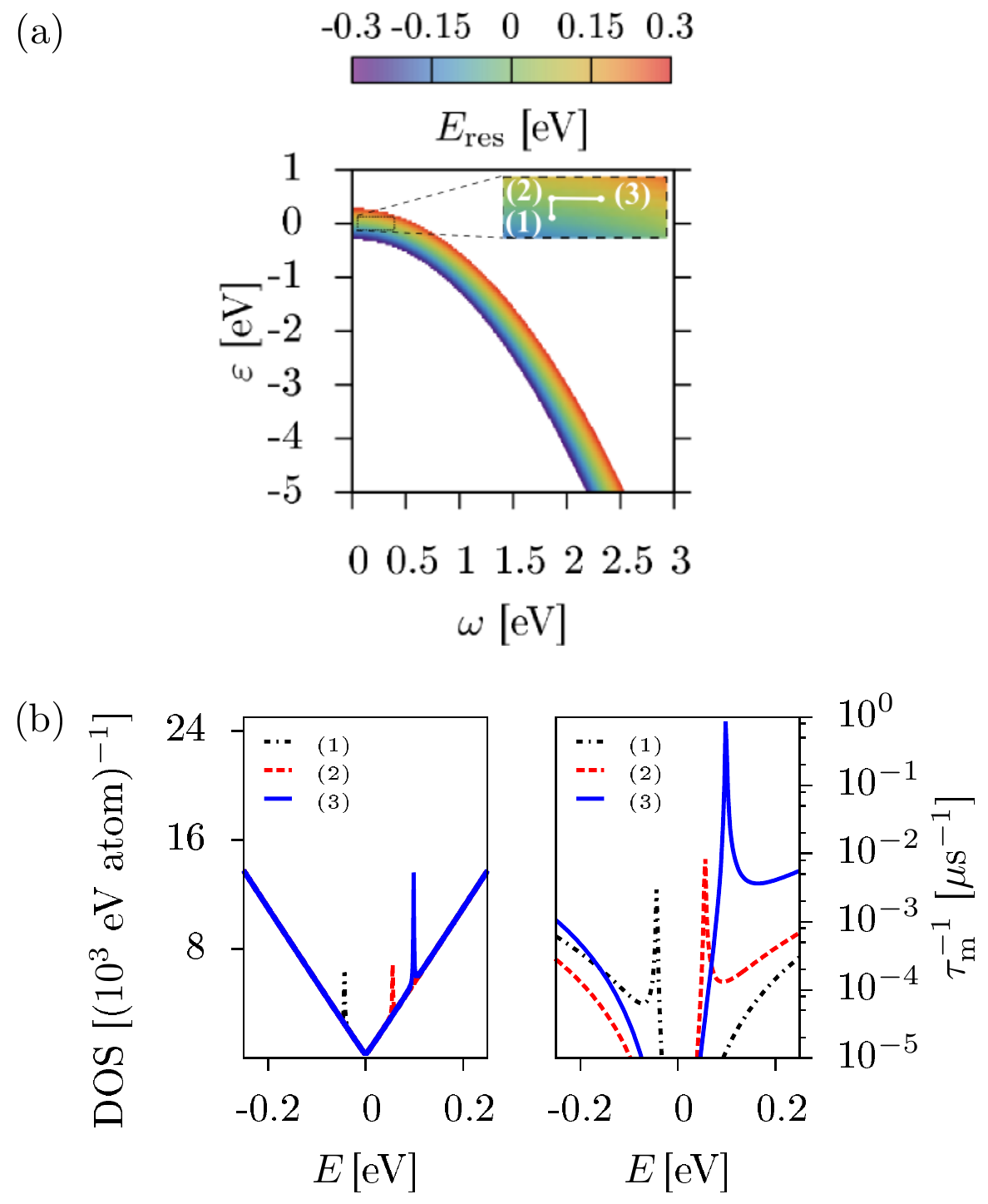}
\caption{\label{Fig:hollowResonanceMap_DosMomrel}(Color online) Resonance and momentum relaxation characteristics due to adatoms in the hollow position. (a) Resonance energy $E_\mathrm{res}$ is shown as a function of $\varepsilon$ and $\omega$. Widths $\Gamma$ (not shown) are not resolved and are estimated to be $\Gamma\leq 2\,$meV. (b) Snapshots of DOS and $\tau_m^{-1}$ at three parameter sets (1) $\omega = 0.2\,$eV and $\varepsilon = -0.08\,$eV, (2) $\omega = 0.2\,$eV and $\varepsilon = 0.02\,$eV, and (3) $\omega = 0.3\,$eV and $\varepsilon = 0.02\,$eV. The resonance levels are, respectively, at (1) $E_{\rm res}=-43\,$meV, (2) $E_{\rm res}=55\,$meV, and (3) $E_{\rm res}=99\,$meV. DOS data is shown for adatom concentration of $\eta = 500\,$ppm, for better resolution, and momentum relaxation rates for realistic $\eta=1\,$ppm.}
\end{figure}

\begin{figure}
{\centering\includegraphics[width=0.9\columnwidth]{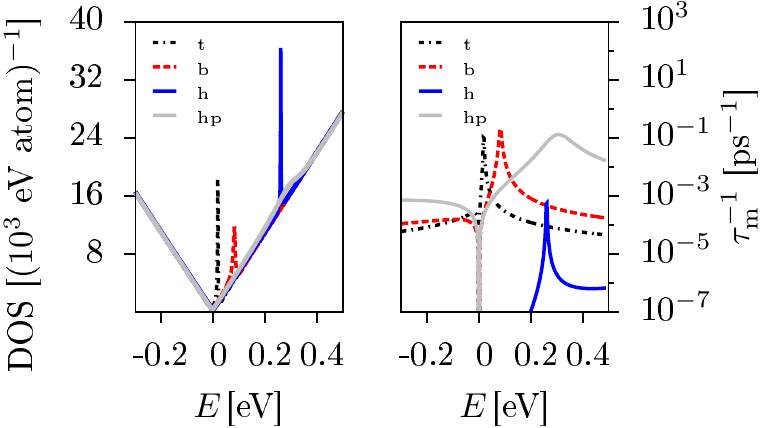}}
\caption{\label{Fig:topbridgehollow_dosmomrel}(Color online) Left and right panel show DOS and momentum relaxation rate, respectively, for an adatom with $\omega = 0.54\,$eV and $\varepsilon = 0.02\,$eV in different adsorption positions, namely top (t), bridge (b), and hollow (h). Corresponding resonant energies and FWHMs are (t) $E_\mathrm{res} = 18\,$meV and $\Gamma = 3\,$meV, (b) $E_\mathrm{res} = 82\,$meV and $\Gamma = 10\,$meV, and (h)~$E_\mathrm{res} = 261\,$meV and $\Gamma \leq 2\,$meV. A toy model calculation for the hollow position (hp) (see main text) shows strong coupling between adatom and graphene compared to (h), leading to a resonance peak at $E_\mathrm{res} = 286\,$meV, $\Gamma=114\,$meV. For better visibility, a concentration of $\eta = 500\,$ppm is used in the DOS, the momentum relaxation rate data is shown for $\eta = 1\,$ppm.}
\end{figure}

An adatom in the hollow position that preserves the $C_{6v}$ symmetry of the hexagonal graphene, as realized in the model of Sec.~\ref{sec:formal}, is affected by destructive interference of electrons tunneling to the adatom.
The effective decoupling of the adatom from a wide range of graphene continuum states in the Brillouin zone \cite{Ruiz2016} leads to distinctive features in several contexts such as local spin-orbit coupling \cite{Weeks2011,Brey2015}, scanning tunneling spectroscopy \cite{Uchoa2009,Saha2010,Wehling:PRB2010,Uchoa2014}, Kondo effect \cite{Uchoa2011,Ruiz2017}, Anderson localization \cite{Garcia2014}, and graphene for chemical sensing \cite{Duffy2016}. We focus here on the dependence of the resonance level on the orbital parameters describing the hollow adatom in direct comparison to the top and bridge adsorption positions and show the results for a large parameter space, which was to our knowledge not addressed before.

Figure~\ref{Fig:hollowResonanceMap_DosMomrel}(a) displays the drastic reduction of the $(\omega,\varepsilon)$-parameter space for resonance levels in energy range~$[-0.3,0.3]\,$eV. Following the path (1)-(3) in parameter space we see similar dependence of the position of resonance levels on $\omega$ and $\varepsilon$ as in bridge position. We do not display corresponding peak widths as we can not resolve them satisfactorily. The restriction to the resolution of the peak widths comes from an energy broadening $\delta=1\,$meV in the calculation of Green's functions that we keep due to numerical reasons (see Appendix\ref{app:green}). We estimate the width $\Gamma\leq 2$\,meV for all addressed peaks in the DOS. 

The destructive interference becomes especially visible in particular DOS and $\tau_m^{-1}$ calculations displayed at Fig.~\ref{Fig:hollowResonanceMap_DosMomrel}(b). The adatom level presents itself as a very sharp peak in the DOS and is strongly sensitive to variation of $\omega$ and $\varepsilon$. This sensitivity is more pronounced than in the bridge case. Even at higher energies where one would expect a stronger hybridization of the impurity state with the graphene due to larger availability of graphene states, the peaks show no broadening but sit on top of the DOS of pristine graphene. Furthermore, we see that the momentum relaxation rates for $E\neq E_{\mathrm{res}}$ are much smaller than for top or bridge for the same $\omega$ and $\varepsilon$. Figure~\ref{Fig:topbridgehollow_dosmomrel} shows a comparison of the DOS and $\tau_m^{-1}$ for the three adsorption positions. The hollow momentum relaxation rate only increases with larger peak energy. The hollow adatom, $m_z=0$, appears therefore as a weak scatterer in graphene compared to top and bridge position. This result is in accordance with the findings of Ref.~\onlinecite{Duffy2016} where the authors investigate the scattering cross section further.

Overall, we find in our analysis of resonance levels, DOS and $\tau_m^{-1}$ clear signatures for the decoupling of the hollow adatom from the graphene continuum states. The decoupled state is also visible in the tight-binding band structure of graphene supercells with hollow adatoms as an dispersionless energy band on top of graphene's band structure (see Appendix~\ref{app:toyModels}). 

We can break the destructive interference, for example, by considering a toy model describing an adatom orbital with magnetic quantum number $m_z=1$ in hollow position (see Appendix~\ref{app:toyModels}). As shown in Fig.~\ref{Fig:topbridgehollow_dosmomrel}, the toy model momentum relaxation rate is now comparable in magnitude to the top and bridge adatom. Broadening of peaks in the DOS as well as the enhanced momentum relaxation rate indicate the lifting of the destructive interference seen for $m_z=0$ and restored effective coupling to the graphene states.

\subsection{\label{sec:local}Localization of resonant states}

\begin{figure}
\includegraphics[width=\columnwidth]{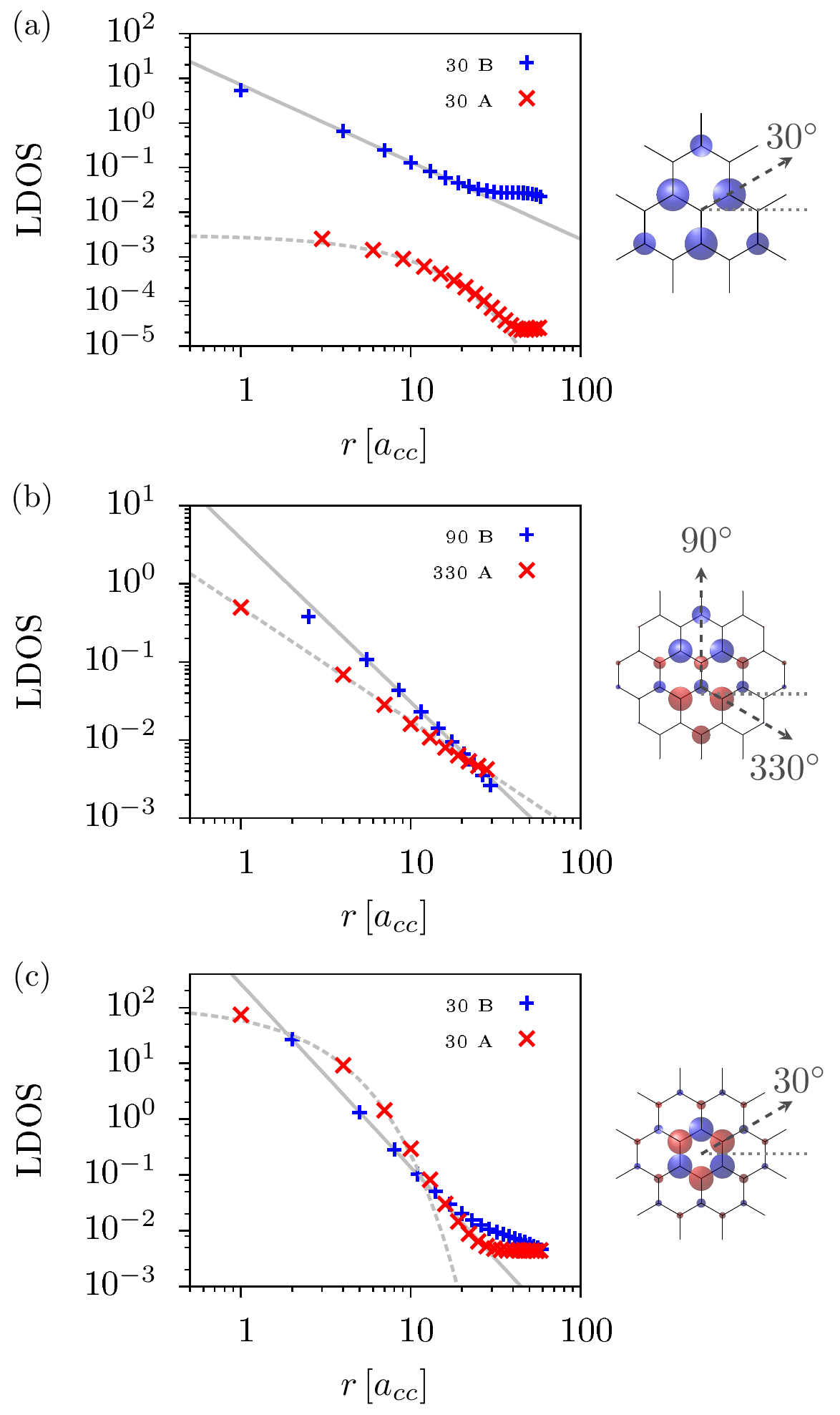}
\caption{\label{Fig:locLDOS}Localization of resonant states in graphene around (a) top, (b) bridge, and (c) hollow adatom along selected directions. Insets show scaled LDOS (sphere radii indicate magnitude) calculated from the $T$-matrix formalism with down-folded Hamiltonian on the graphene lattice. The colors refer to the two sublattices A (red) and B (blue). Top and bridge position show power-law decay in the dominant directions (see parameters in the main text), whereas hollow shows no clear tendency.}
\end{figure}

To investigate the localization of states around the impurity at resonance energy, we calculate with the triangle method \cite{Kurganskii1985} the local density of states (LDOS) for representative adatoms from tight-binding supercells of size 40$\times$40. These supercells mimic a dilute adatom concentration of about $312\,$ppm. We extract the LDOS at the peak energy $E^{\rm tb}_{\rm res}$ of the DOS which is found in all cases close to the predicted resonance energies $E_{\rm res}$ for the single adatom limit. Figure~\ref{Fig:locLDOS} displays the LDOS dependence on the distance from the adatom along selected directions in the case of a specific top, bridge, and hollow adatom.

First, we note that the LDOS around top, bridge, and hollow positions shows symmetrical behavior in accordance with the local point group symmetries $C_{3v}$, $C_{2v}$, and $C_{6v}$, respectively, of graphene around the corresponding adatom. This symmetry is also seen in the LDOS as calculated from the down-folded Hamiltonian in the $T$-matrix formalism (see Fig.~\ref{Fig:locLDOS}). The tight-binding supercell calculations are used for the quantitative analysis. Due to the finite size of the supercells we can only investigate the short-range behavior around the impurity---approaching the supercell border, the LDOS values saturate.

The wave function profile for the top position was, for example, already investigated in Ref.~\onlinecite{Liang:PRL2012} in the single adatom limit where it was also pointed out that the power-law decay exponent depends on the direction of the path taken away from the impurity and the sublattice of the investigated site. Furthermore, it is well known that for a strong scatterer in the top position the resonant state is more localized on the opposite sublattice \cite{Farjam2011}---in the extreme case of a vacancy the resonant state populates exclusively the opposite (intact) sublattice and the LDOS decays \cite{Pereira2006,Pereira2008,Nanda2012} as $r^{-2}$. For top adatoms, the decay exponents depend on the orbital parameters of the chosen adatom and thus its resonance energy. For $\omega=0.54\,$eV and $\varepsilon=0.02\,$eV, corresponding $E^{\rm tb}_{\rm res}=19\,$meV, the LDOS is significantly smaller on sublattice A to which the adatom is adsorbed than on the opposite sublattice B, see Fig.~\ref{Fig:locLDOS}(a). Along a selected line, starting at the adsorption position with $30^{\circ}$ with respect to the $x$-axis, we extract a power-law decay on sublattice B, $|\psi|^2 \propto r^{-p}$, $p\simeq 1.72$, and a tendency to exponential decay, $|\psi|^2 \propto \exp{(-q r)}$, $q\approx 0.12$, on sublattice A.
The larger the resonant energy the broader the resonance peak gets due to stronger interaction with the graphene states, and the contributions on both sublattices will approach each other \cite{Liang:PRL2012}.

The LDOS distribution for the bridge position around the adatom is shown in Fig.~\ref{Fig:locLDOS}(b). Its diamondlike shape reminds of two intertwined triangles with their centers on neighboring sites A and B. This appearance looks natural if one imagines the bridge adatom as two neighboring top adatoms or double substitutional impurities \cite{Wehling2007}. The LDOS contribution on the hybridization partners of the bridge adatom is decreased due to the effective coupling between them which is mediated by the adatom [see Eq.~(\ref{Eq:GeneralAdatomPert})]. The overall resulting population of the two sublattices around the bridge adatom shows power-law decay: For a copper adatom on graphene in the bridge position \cite{Frank:PRB2017}, $\omega=0.54\,$eV and $\varepsilon=0.02\,$eV, energy $E^{\rm tb}_{\rm res}=83\,$meV, we select two directions along $90^\circ$ and $330^\circ$ shown in Fig.~\ref{Fig:locLDOS}(b) with a clear power-law decay with $p\approx 2.09$ and $p\approx 1.45$, respectively.

As in the bridge case, the hollow adatom does not distinguish sublattices which is clearly seen in the LDOS distribution in Fig.~\ref{Fig:locLDOS}(c). Extracting the LDOS on the lattice sites along the $30^\circ$ direction for a hollow adatom with $\omega=0.3\,$eV, $\varepsilon=0.02\,$eV and peak energy $E^{\rm tb}_{\rm res}=99\,$meV we see tendencies to both power-law and exponential decay, $p\approx 3.29$ and $d\approx 0.62$ depending on the sublattice. Note that the resonance energy is comparable to the previous bridge example. We know from the previous section that the hollow position ($m_z=0$) suffers from destructive interference which seems to be related to the occurrence of both power-law and exponential behavior on the same order of magnitude. Indeed, we found that with lowering the adatom content in supercell calculations the hollow adatom ($m_z=0$) loses LDOS contribution much faster than the top or bridge adatom. On the contrary, a clear power-law decay is seen for a toy calculation with $m_z=1$ for the same orbital parameters (see Appendix~\ref{app:toyModels}, Fig.~\ref{Fig:hollow6ResonanceMap_DosMomrel}(c)).

\subsection{\label{sec:vac}Vacancy vs. adatom}

\begin{figure}
  {\centering\includegraphics[width=0.95\columnwidth]{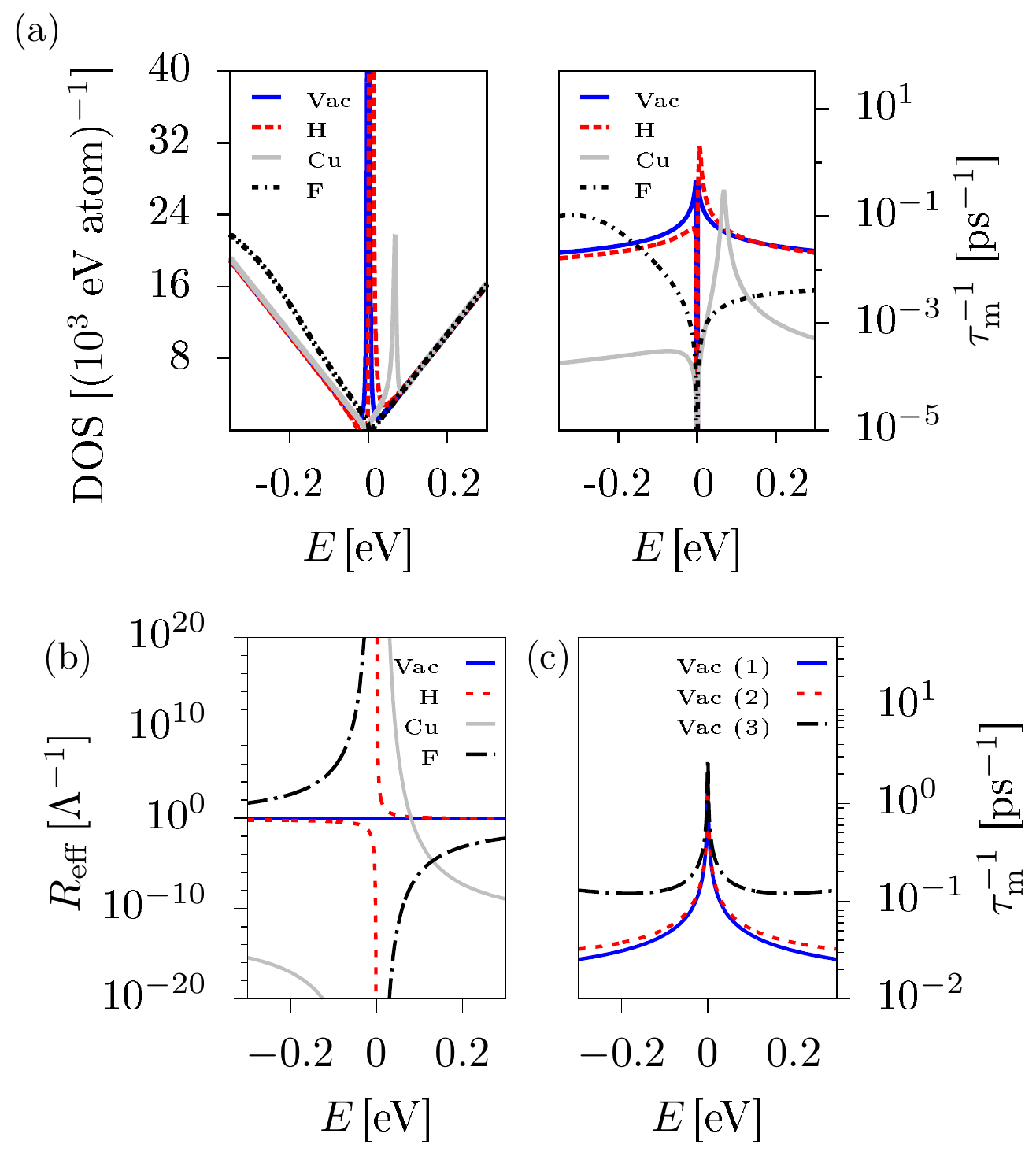}}
  \caption{\label{Fig:topVacMomrel}(Color online) Adatom in top position and vacancy model. Panel (a) compares DOS (left) and momentum relaxation rate (right), respectively, of a vacancy (Vac) with adatoms in the top position: hydrogen adatom (H), $\omega = 7.5\,$eV and $\varepsilon~= 0.16\,$eV, copper adatom (Cu), $\omega = 0.81\,$eV and $\varepsilon = 0.08\,$eV, and fluorine (F), $\omega = 5.5\,$eV and $\varepsilon = -2.2\,$eV. For better resolution, a concentration of $\eta = 10^{3}\,$ppm is used in the DOS, the momentum relaxation rate data are shown for $\eta = 1\,$ppm. Fluorine in the top position induces a broad resonance, $\Gamma=277\,$meV, at about $E_\mathrm{res}=-262\,$meV, leading to a small shoulder in the DOS and momentum relaxation rate at negative energies. Copper and hydrogen show stronger features at lower energies: copper induces a resonance level at $E_\mathrm{res}=68\,$meV with $\Gamma=9\,$meV. Hydrogen with $E_\mathrm{res}=7\,$meV, $\Gamma=5\,$meV, is comparable to a vacancy at zero energy, which is fully symmetric with respect to negative and positive energies. Panel (b) shows the dependence of the effective impurity radius $R_{\mathrm{eff}}$ on the energy for a vacancy as well as hydrogen, copper, and fluorine adatoms. Panel (c) displays the momentum relaxation rate for a vacancy under different approximations to Eq.~(\ref{Eq:cond_reff}). Neglecting the imaginary part of $G_{00}$, graph (2), increases $\tau_m^{-1}$ slightly (by about $20\%$ at $E=200\,$meV) compared to the exact result, graph (1). Further overestimation of $\tau_{m}^{-1}$ (by a factor of 4 at $E=200\,$meV) originates in an artificially increased $R_{\mathrm{eff}}=4.5\Lambda^{-1}$, graph (3).}
\end{figure}

The top-position adatom is often compared to a vacancy, the prototype of a strong resonant scatterer. This structural defect in graphene induces a sharp midgap state at zero energy. Leaving aside reconstruction \cite{El-Barbary2003,Amara2007} of a single vacancy site in graphene, the vacancy can be modelled either by removing the vacancy site from the graphene lattice or, equivalently, by assigning a local potential $U$ to the corresponding site and taking $U\to\infty$ \cite{Peres:PRB2006,Pereira2008,Ducastelle:PRB2013}. In our model description of the top adatom, this would mean to send the effective on-site potential $\omega^2/(E-\varepsilon)\to\infty$ in Eq.~(\ref{Eq:tmatGeneral}). We obtain the $T$-matrix for a vacancy \cite{Peres:PRB2006,Ferreira:PRB2011},
\begin{equation}
\mathcal{T}_{vac} = -\frac{1}{G_{00}(E)} \: \ket{c_0}\bra{c_0}\:. \label{Eq:tmatVac}
\end{equation}
The DOS and $\tau_m^{-1}$ resulting from Eq.~(\ref{Eq:tmatVac}) are shown in Fig.~\ref{Fig:topVacMomrel}(a). The vacancy introduces a sharp resonant peak at zero energy, fully symmetric with respect to negative and positive energies.
This symmetric appearance does not hold for general top adatoms as presented in Section~\ref{sec:top}. DOS and $\tau_m^{-1}$ for the adatoms fluorine \cite{Irmer:PRB2015}, copper \cite{Frank:PRB2017}, and hydrogen \cite{Gmitra:PRL2013} are included in Fig.~\ref{Fig:topVacMomrel}(a). The asymmetry is very small for a strong resonant scatterer in the top position, such as hydrogen with a resonance level close to zero energy, which leads to the similarity of a hydrogen adatom to a vacancy in graphene.

Using the Boltzmann transport formalism, with the transition rates from the $T$-matrix and the analytic result for the on-site Green's function $G_{00}$ (Appendix~\ref{app:green}), we obtain the conductivity for graphene in the presence of an adatom in the top position,
\begin{equation}\label{Eq:cond_reff}
  \sigma = \frac{e^2}{h} \frac{4}{\pi\eta}\frac{E^2}{W^2} \left[\ln^2\left(\frac{|E|}{\hbar v_F} R_{\mathrm{eff}}\right) + \frac{\pi^2}{4}\right]\:,
\end{equation}
where we have introduced $R_{\mathrm{eff}}(E)$,
\begin{equation}\label{Eq:reff}
  R_{\mathrm{eff}}(E) = \Lambda^{-1} \exp{\left[-\frac{1}{2}\frac{W^2}{\omega^2}\frac{(E-\varepsilon)}{E}\right]}\:.
\end{equation}
with the momentum cut-off $\Lambda$ (see Appendix~\ref{app:green}). The quantity $R_{\mathrm{eff}}(E)$ has the dimension of a length and can be interpreted as an effective radius of a top positioned scatterer. For a vacancy we get $R^{\mathrm{Vac}}_{\mathrm{eff}} = \Lambda^{-1}\approx 0.9\,${\AA}. Figure~\ref{Fig:topVacMomrel}(b) displays the energy dependence of $R_{\mathrm{eff}}$ for the top adatoms hydrogen, copper, and fluorine, in comparison to a vacancy. The effective radius diverges at zero energy for the adsorbates.

We can directly compare Eq.~(\ref{Eq:cond_reff}) for a top adatom to the model of a \emph{strong midgap scatterer} (SMS) or vacancy of Refs.~\onlinecite{Stauber2007,Katsnelson2007,Ferreira:PRB2011}. There, the defect is modeled as a potential disk with finite (energy independent) radius $R$. The scattering cross section and conductivity are obtained from partial wave decomposition. The conductivity reduces to \cite{Stauber2007,Katsnelson2007,Ferreira:PRB2011}
\begin{align}
  \sigma_{\mathrm{SMS}} &= \frac{e^2}{h} \frac{2 k^2_{F}}{\pi^2 n_i} \ln^2(k_F R)\nonumber\\
  &= \frac{e^2}{h} \frac{2}{\pi} \frac{2}{A_{\mathrm{uc}}n_i}\frac{E^2}{W^2} \ln^2\left(\frac{|E|}{\hbar v_F} R\right)\:,\label{Eq:cond_SMS}
\end{align}
where $n_i$ is the impurity concentration per unit area \cite{Ferreira:PRB2011}. Comparing the result of the SMS model to Eq.~(\ref{Eq:cond_reff}), we obtain $\sigma=2\cdot\sigma_{\mathrm{SMS}}$ if we set $R_{\mathrm{eff}}=R$ and neglect the term $\pi^2/4$ which originates from the imaginary part of $G_{00}$. The quantities $\eta$ and $n_i$ are related by $\eta = 2 n_i/A_{\mathrm{uc}}$ where $A_{\mathrm{uc}}$ is the unit cell area. The additional factor of two was also found in the vacancy study of Ref.~\onlinecite{Wehling:PRL2010}.

Note that our $T$-matrix formulation uses a fixed momentum cut-off that preserves the number of states. In the analysis of experimental data one uses $\sigma_{\mathrm{SMS}}$ and fits the radius $R$ together with $n_i$, assuming that $R$ is at the order of a few angstroms \cite{Chen2009,Ni2010,Hong:PRL2012}. Figure~\ref{Fig:topVacMomrel}(c) shows the effect of the approximations to the momentum relaxation rate for a vacancy. Neglecting the imaginary part of $G_{00}$ and using $R=4.5 \cdot \Lambda^{-1}\approx 4.1\,${\AA}, the momentum relaxation rate at $E=200\,$meV for a vacancy is overestimated by a factor of four. As the momentum relaxation rate is directly proportional to the impurity concentration, a simultaneous fit of $n_i$ and $R$ to experimental resistivity data can lead to an underestimation of $n_i$.

Clearly, the model for strong midgap scatterers is not designed to reflect adatoms with resonance energies significantly different from zero. The previous sections have shown that the resonance level strongly depends on the orbital parameters and equally on the adsorption position. Fluorine, an adatom also binding in the top position \cite{Irmer:PRB2015}, induces a dominant asymmetry in the momentum-relaxation rate, Fig.~\ref{Fig:topVacMomrel}(a), due to a far-off (at about -300 meV) and broad resonance. A single fluorine adatom or dilute concentration of non-interacting fluorine adatoms is not captured by a vacancy or SMS model. Note, that the SMS model yields generally higher $\tau_m^{-1}$ than the $T$-matrix model for adsorbates, which can be understood as the consequence of several approximations to $G_{00}$ in the SMS approach. In experiments where the conductivity is well described by Eq.~(\ref{Eq:cond_SMS}), also additional sources of strong midgap scatterers, charged impurities \cite{Adam2007}, or clusters \cite{McCreary2010,Katsnelson2009} can play a role so that the analysis of the experimental data has to be done more carefully.

\section{\label{sec:sum}Summary}
In summary, we studied the effect of adsorption position and orbital parameters (on-site energy $\varepsilon$ and hybridization strength $\omega$) of single adatoms on graphene to the formation of resonances in an energy range that is accessible by experiments. Overall, we find significant differences between the three adsorption positions top, bridge, and hollow.

In the top position, the resonance level lies closer to zero energy the larger $\omega$, or the smaller $\varepsilon$, is. Especially, the resonance energy $E_{\rm res}$ changes sign under $\varepsilon \to -\varepsilon$. The resonance levels leave distinct features in DOS and $\tau_m^{-1}$. For resonance energies far away from zero energy, a pronounced electron-hole asymmetry is predicted.

For the bridge position, we find that the resonance level is more sensitive to changes in orbital parameters: the parameter range leading to resonances in the studied energy range is dominated by negative $\varepsilon$ and increasing $\omega$ shifts resonance levels to higher energies. For same orbital parameters, the resonance level for the bridge position lies at higher energy than for the top position. Rates $\tau_m^{-1}$ are in magnitude comparable to the top position.

A hollow adatom with $s$ or $p_z$ orbital, on the contrary, acts as a weak scatterer in graphene as it is effectively decoupled from graphene due to destructive interference of electrons hopping on and off the adatom. Resonance levels are seen within the studied energy range only for a narrow window of orbital parameters. Furthermore, the resonance peaks resulting in DOS and $\tau_m^{-1}$ are very narrow, and the rates $\tau_m^{-1}$ are significantly smaller than in the top or bridge positions. From the LDOS calculation of tight-binding supercells, we find that the resonance state induced by a hollow adatom shows no clear tendency to power-law localization. On the contrary, top and bridge adatoms give rise to resonant states with a clear power-law decay.

Finally, we compare our findings valid for monovalent adatoms on graphene with vacancies and the SMS model of Refs.~\onlinecite{Stauber2007,Katsnelson2007,Ferreira:PRB2011}. Both vacancy and SMS induce zero energy resonances resulting in electron-hole symmetric $\tau_m^{-1}$, which is also approximately the case for adatoms with a resonance level very close to zero energy, for example, hydrogen \cite{Gmitra:PRL2013}. However, variation of the effective defect radius $R$ in the SMS model can enhance (and overestimate) $\tau_m^{-1}$ significantly compared to vacancies or single adatoms from our $T$-matrix analysis. Therefore analyzing experimental data for adsorbates on graphene one has to take carefully into account the particular limits of the different models.

Our results can help to understand experimental transport studies with dilute adatom concentrations on graphene. Especially our comprehensive resonance maps, i.e., the dependence of scattering on different adatoms in different adsorption positions, can help to clarify the role of specific adatoms in the limit of short-range scattering.

\begin{acknowledgments}
This work was supported by the European Union's Horizon 2020 research and innovation programme under grant agreement No. 696656, the DFG SFB Grants. No. 689 and 1277 (A09), and GRK Grant No. 1570.
\end{acknowledgments}

\appendix

\section{\label{app:green}Green's functions for graphene}

\begin{figure}
\includegraphics[width=0.35\columnwidth]{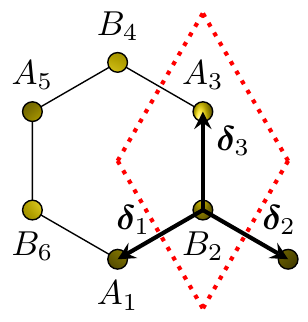}
\caption{\label{Fig:defCellHexagon}(Color online) Site labeling convention inside the graphene unit cell (red dashed diamond) as used for the real space representation of the Green's functions. The nearest-neighbor connection vectors $\mathbf{\delta}_j$, $j = 1,2,3$ are displayed by bold arrows.}
\end{figure}

Since we investigate resonant scattering with local impurities, we focus on the real space representation of retarded Green's functions. We start from the (full) $k$-dependent Hamiltonian for graphene whose eigenenergies are given by
\begin{equation} 
\varepsilon_{nk} = nt \left|f(k)\right|,
\end{equation}
where $n=1$ for the conduction band and $n=-1$ for the valence band and $f(k)=\exp(\mathrm{i} \mathbf{k}.\boldsymbol{\delta}_1) + \exp(\mathrm{i} \mathbf{k}.\boldsymbol{\delta}_2) + \exp(\mathrm{i} \mathbf{k}.\boldsymbol{\delta}_3)$. See Fig.~\ref{Fig:defCellHexagon} for our convention of the unit cell and the nearest-neighbor vectors $\boldsymbol{\delta}_{j}=(\cos(j\frac{2\pi}{3}-\frac{\pi}{6}), \sin(j\frac{2\pi}{3}-\frac{\pi}{6})) a_{cc}$ with $j=1,2,3$.

Inverting Eq.~(\ref{Eq:greensDef}) and performing the Fourier transformation to real space, we obtain three different kinds of Green's functions: first, the on-site Green's function,
\begin{equation}
G_{00}(E)=\lim_{\delta \to 0} G_{00}(E^+)=\lim_{\delta \to 0} \frac{V}{N} \int \frac{\mathrm{d}k^2}{(2\pi)^2} \frac{E^+}{{E^+}^2-t|f(k)|^2} \:,\label{Eq:GreenOn}
\end{equation}
second, the Green's function coupling opposite sublattices, 
\begin{align}
G^{\rm AB}_{lm}(E) &=\lim_{\delta \to 0} G^{\rm AB}_{lm}(E^+)\nonumber\\
&=\lim_{\delta \to 0} \frac{V}{N} \int \frac{\mathrm{d}k^2}{(2\pi)^2} \frac{-t f(k) \mathrm{e}^{\ii \mathbf{k}.(\mathbf{R}_l-\mathbf{R}_m)}}{{E^+}^2-t|f(k)|^2} \:, \label{Eq:GreenOpp}
\end{align}
describing the propagation from site $\mathbf{R}_m$ on sublattice $B$ to site $\mathbf{R}_l$ on sublattice $A$, and, third,
the Green's function coupling different sites at $\mathbf{R}_m$ and $\mathbf{R}_l$ on the same sublattice $\mathrm{N}=\mathrm{A},\mathrm{B}$,
\begin{align}
G^{\rm NN}_{lm}(E) &=\lim_{\delta \to 0} G^{\rm NN}_{lm}(E^+)\nonumber\\
&=\lim_{\delta \to 0} \frac{V}{N} \int \frac{\mathrm{d}k^2}{(2\pi)^2} \frac{E^+ \mathrm{e}^{\ii \mathbf{k}.(\mathbf{R}_l-\mathbf{R}_m)}}{{E^+}^2-t|f(k)|^2} \:.\label{Eq:GreenSame}
\end{align}
Interchange of sublattices leads to complex conjugation of $f(k)$ in Eq.~(\ref{Eq:GreenOpp}). Note that in this convention, vectors  $\mathbf{R}_m$ point to the actual positions of the lattice sites $m$ and not to the unit cell hosting those sites. For the calculation of the DOS, see main text Sec.~\ref{sec:formal}, we use the derivative of the Green's operator, $\frac{\partial}{\partial E} \mathcal{G}_0$, which we obtain in real-space from Eqs.~(\ref{Eq:GreenOn})-(\ref{Eq:GreenSame}) by differentiating the integrand with respect to energy $E$.

There are several methods to calculate the Green's functions for graphene.
One, rather intuitive, approach relies on the linear approximation of the spectrum around the $k$-points $\mathbf{K}^{\pm}$ in the first Brillouin zone. The integration is transformed to two integrals around $\mathbf{K}^{\pm}$ up to a momentum cutoff $\Lambda$, which ensures conservation of the number of states. An analytic result can then be directly obtained for $G_{00}$,
\begin{equation}\label{Eq:G00}
G_{00}(E) = \frac{E}{W^2}\ln\left|\frac{E^2}{W^2-E^2}\right| - \ii \pi\frac{|E|}{W^2} \Theta(W-|E|),
\end{equation}
where $W=\hbar v_F \Lambda = \sqrt{\sqrt{3}\pi}t$ is the cut-off energy and $\Theta(x)$ the Heaviside step function.

On the contrary, analytic formulas for $G^{\rm AB}_{lm}$ and $G^{\rm NN}_{lm}$ require the approximation $\Lambda\to\infty$ \cite{Ducastelle:PRB2013,Bundesmann:PRB2015}.
As already pointed out by Refs.~\onlinecite{Ducastelle:PRB2013,Ihnatsenka:PRB2011}, this approximation should be taken with a grain of salt and the momentum-cutoff has to be chosen carefully in general \cite{Wang:PRB2006}.

In fact, we found that applying $\Lambda\to\infty$ in $G^{\rm AB}_{lm}$ and $G^{\rm NN}_{lm}$ has severe effects on the resonance energy calculation in the bridge and hollow positions of Secs.~\ref{sec:bridge} and \ref{sec:hollow}. Resonance levels in the bridge position shift significantly: the resonance level of copper in the bridge position, as calculated in Ref.~\onlinecite{Frank:PRB2017}, changes from $E_{\rm res}=128\,$meV to $E_{\rm res}=82\,$meV when going from the linear approximation with $\Lambda\to\infty$ to numeric integration over the full Brillouin zone. Furthermore, peaks observed in tight-binding supercell calculations for hollow adatoms are absent in the calculation with the linearized model and $\Lambda\to\infty$.

We therefore use full numerical integration over the 2D Brillouin zone to obtain all Green's functions that we need for the calculations discussed in the main text. Due to computational reasons we keep a finite imaginary part of $\delta = 10^{-3}\,$eV in our calculations which induces energy broadening. We checked, by rescaling $\delta\to\delta/2$, that the finite imaginary part only marginally affects the resonance position (at the order of 0.1\%). Naturally, the finite energy-broadening affects the widths of the resonance peaks. In the top and bridge positions, the widths $\Gamma$ decrease by about 1 to 2\,meV upon $\delta\to\delta/2$. Very narrow peaks $\Gamma<2\,$meV can not be resolved for $\delta=1\,$meV. Therefore, in the hollow position, for $m_z=0$, all peaks widths appear to be twice the energy broadening. In the top and bridge positions, we obtain, despite finite $\delta$, the correct order of magnitude for $\Gamma$ values.

Symmetry considerations reduce the number of Green's functions that are needed for the different adsorption positions. Green's functions $G^{\rm AB}_{lm}$ show, apart from the natural translational symmetry, threefold rotational symmetry. Green's functions $G^{\rm NN}_{lm}$ are invariant under sixfold rotations \cite{Bacsi:PRB2010}. Furthermore, it holds that $G^{\rm AB}_{lm}(E)=G^{\rm BA}_{ml}(E)$.
We thus end up with a maximum of four Green's functions (and corresponding derivatives) that have to be evaluated for resonance energy calculations in the hollow position, $G_{00}$, $G^{\rm AB}_{12}=G^{\rm AB}_{16}=G^{\rm AB}_{32}$, $G^{\rm NN}_{13}=G^{\rm NN}_{15}=G^{\rm NN}_{35}$ and $G^{\rm AB}_{14}=G^{\rm AB}_{52}=G^{\rm AB}_{36}$. For our convention of the site labeling see Fig.~\ref{Fig:defCellHexagon}.

\section{\label{app:momrel}Momentum relaxation rates}

We consider elastic scattering off adatoms on graphene, excluding multiple-scattering events which are suppressed in the dilute (single) adatom limit.
The starting point is the transition rate between states $\ket{n k}$ and $\ket{n^\prime k^\prime}$ of same eigenenergy $\varepsilon_{nk}$, expressed by the generalized Fermi's golden rule,
\begin{equation}
W_{nk|n^\prime k^\prime} = \frac{2\pi}{\hbar} \left|\bra{n^\prime k^\prime }  \mathcal{T}(\varepsilon_{nk}) \ket{n k}\right|^2 \delta(\varepsilon_{nk}-\varepsilon_{n^\prime k^\prime})\:.
\end{equation}
Symbol $n=\pm 1$ denotes conduction and valence band, and $\mathbf{k}$ is the wave vector. The adatom orbital parameters and adsorption position enter the rate via the $T$-matrix term.

From the transition rate we obtain (for isotropic scattering in $k$ space) the elastic transport scattering rate by weighting the rate with the transport factor $(1-\cos \phi_{kk^\prime})$ and summing over final states, i.e.,
\begin{equation}
{\tau_{nk}}^{-1} = \sum\limits_{k^\prime ,n^\prime} (1-\cos \phi_{kk^\prime})\:  W_{nk|n^\prime k^\prime}\,.
\end{equation}
Here,$\phi_{kk^\prime}$ is the angle between $\mathbf{k}$ and $\mathbf{k^\prime}$ vectors.
Averaging over the Fermi-contour gives the momentum relaxation rate at given energy $E$,
\begin{equation}
\tau^{-1}(E) = \frac{\sum\limits_{k ,n} {\tau_{nk}}^{-1} \delta(E-\varepsilon_{nk})}{\sum\limits_{k ,n} \delta(E-\varepsilon_{nk})}\:. 
\end{equation}
As we are interested in the effects near the charge neutrality point we use the linearized graphene spectrum around the Dirac points.. The momentum relaxation rates are directly proportional to the adatom concentration $\eta=1/(2N)$.

\section{\label{app:toyModels}Hollow adatom - toy model}

\begin{figure}
\includegraphics[width=\columnwidth]{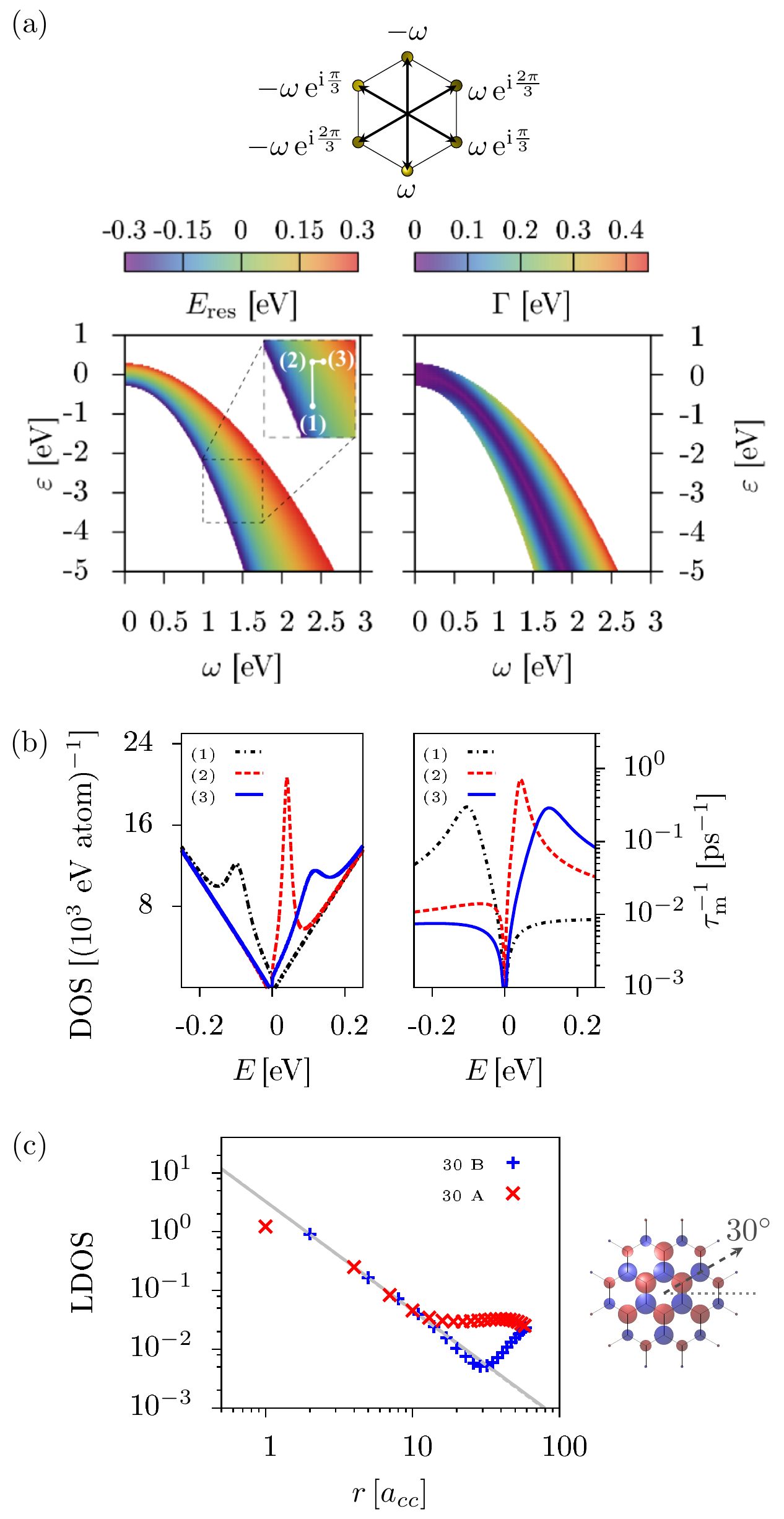}
\caption{\label{Fig:hollow6ResonanceMap_DosMomrel}(Color online) Resonance and momentum relaxation characteristics for a toy-adatom in the hollow position ($m_z=1$ orbital). (a) Graphical representation of the phase-dependent coupling between the $m_z=1$ orbital and its carbon hybridization partners as implemented in the model. Resonance energy $E_{\rm res}$ and width $\Gamma$ are shown as functions of $\varepsilon$ and $\omega$. (b) Snapshots of DOS and $\tau_m^{-1}$ at three parameter sets (1) $\omega = 1.4\,$eV and $\varepsilon = -3.3\,$eV, (2) $\omega = 1.4\,$eV and $\varepsilon = -2.5\,$eV, and (3) $\omega = 1.5\,$eV and $\varepsilon = -2.5\,$eV. The resonance levels are at (1) $E_{\rm res}=-94\,$meV with $\Gamma=72\,$meV, (2) $E_{\rm res}=40.3\,$meV with $\Gamma=28\,$meV, and (3) $E_{\rm res}=106.7\,$meV with $\Gamma=92\,$meV. DOS data are shown for adatom concentration of $\eta = 500\,$ppm and momentum relaxation rates for $\eta = 1\,$ppm. (c) LDOS around the impurity and along the $30^{\circ}$ direction are shown for $\omega=0.3\,$eV, $\varepsilon=0.02\,$eV, extracted at $E_{\rm res}=129\,$meV of a $40\times40$ supercell calculation. Both A (red) and B (blue) sublattice contributions follow power-law with a similar exponent.}
\end{figure}

\begin{figure}
\includegraphics[width=0.96\columnwidth]{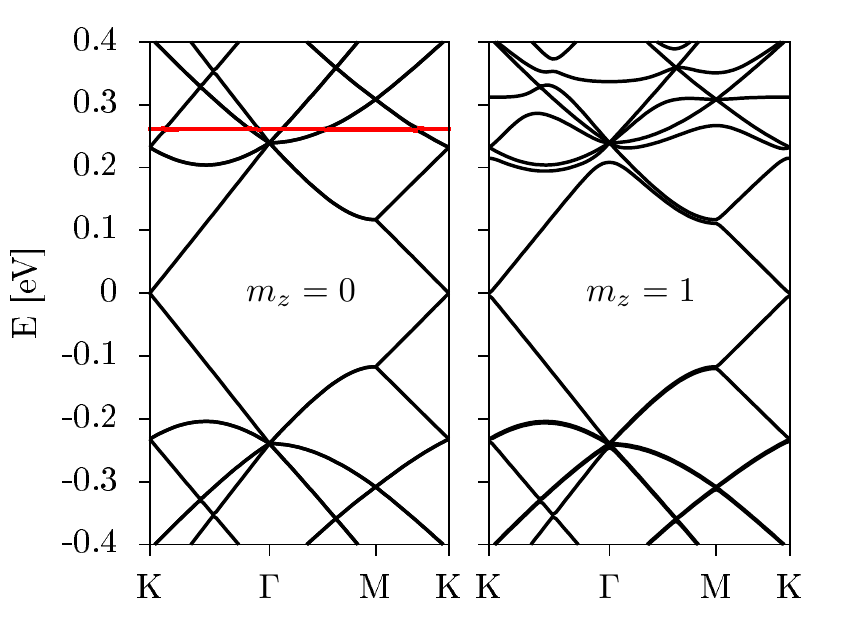}
\caption{\label{Fig:holBands}(Color online) Tight-binding band structure graphene with adatoms in the hollow position with $m_z=0$ (left) and $m_z=1$ (right) orbital for a $40\times 40$ supercell. A flat non-dispersive band (black-red dotted) is observed at about $E=261\,$meV for $m_z=0$ with an almost undisturbed graphene band structure. In the case of $m_z=1$ clear band anti-crossings appear in the energy range $[0.2,0.4]$\,eV showing a strong hybridization of the impurity state with graphene. }
\end{figure}

As seen in Sec.~\ref{sec:hollow}, the adatom with a $m_z=0$ orbital in the hollow position suffers from destructive interference. This is because the adatom orbital couples equally to its six carbon neighbors \cite{Ruiz2016}. Considering an adatom orbital with non-zero magnetic quantum number, the hybridization coupling $\omega$ acquires a different phase for different neighbors and hence the destructive interference can be lifted.

As a toy model, we therefore take into account an atomic orbital with magnetic quantum number $m_z=1$ coupled to all six neighboring carbon sites. Due to transformation of the orbital under the angular momentum operator $L_z$, the hybridization coupling $\omega$ gains a phase factor of $\exp(\mathrm{i} \phi)$ under rotation of angle $\phi$. The modified Hamiltonian $H'$ reads
\begin{equation}
H'=\varepsilon\ket{X}\bra{X}+\sum\limits_{n=1}^{6} \omega_{n} \left(\ket{X}\bra{c_n}+\mathrm{H.c.}\right)+\mathcal{H}_0\:.\label{Eq:HollowAdatomHmod}
\end{equation}
where $\omega_{n} = \omega \exp(\mathrm{i} (n-1) \phi)$ with $\phi = \pi/6 $. Time-reversal symmetry determines whether the coupling $\omega$ is purely real or imaginary. For the present analysis this does not play a role as the hybridization enters the perturbation, Eq.~(\ref{Eq:GeneralAdatomPert}), as the square of the absolute value.

The extension of the model changes significantly the resonances under variation of the orbital parameters $\omega$ and $\varepsilon$, see Fig.~\ref{Fig:hollow6ResonanceMap_DosMomrel}(a). Compared to the $m_z=0$ case, Sec.~\ref{sec:hollow}, a much wider parameter range leads to resonances in the energy interval $[-0.3,0.3]$\,eV, which are also significantly broadened. This is a fingerprint of a strong coupling between adatom and graphene. Band structure calculations for a 40$\times$40 supercell, for $\omega = 0.54$\,eV and $\varepsilon=0.02$\,eV, reveal in the case $m_z=0$ the decoupling of the impurity level from the graphene, see Fig.~\ref{Fig:holBands}. The flat non-dispersive band at about $E=261\,$meV shows that the hybridization with graphene is strongly suppressed. On the contrary, the $m_z=1$ case with the same orbital parameters shows strong coupling between the impurity and graphene with occurrence of a band anti-crossing around the predicted resonance level $E_{\rm res}=285$\,meV ($\Gamma=114$\,meV).

The momentum-relaxation rate, see Fig.~\ref{Fig:hollow6ResonanceMap_DosMomrel}(b), reaches same values as in the top or bridge positions, see Secs.~\ref{sec:top} and \ref{sec:bridge}. Figure~\ref{Fig:hollow6ResonanceMap_DosMomrel}(c) displays the localization of the resonant state in a 40$\times$40 supercell for $\omega=0.3\,$eV and $\varepsilon=0.02\,$eV at energy $E=129\,$meV. Both on A and B sublattice the LDOS decay follows power-law $r^{-b}$ with exponent $b\approx 3.3$.

\bibliography{paper}

\end{document}